\documentclass[pra,aps,twocolumn,showpacs,superscriptaddress,amsmath,amssymb,superscriptaddress]{revtex4-1}
\usepackage{graphicx}
\input{epsf}
\begin{document}
\epsfverbosetrue
\def\la{{\langle}}
\def\ra{{\rangle}}
\def\vep{{\varepsilon}}
\newcommand{\beq}{\begin{equation}}
\newcommand{\eeq}{\end{equation}}
\newcommand{\beqa}{\begin{eqnarray}}
\newcommand{\eeqa}{\end{eqnarray}}
\newcommand{\da}{^\dagger}
\newcommand{\wh}{\widehat}
\newcommand{\os}[1]{#1_{\hbox{\scriptsize{osc}}}}
\newcommand{\cn}[1]{#1_{\hbox{\scriptsize{con}}}}
\newcommand{\sy}[1]{#1_{\hbox{\scriptsize{sys}}}}
\newcommand{\BEC}{}
\newcommand{\q}{\quad}
\newcommand{\lm}{\lambda}
\newcommand{\lt}{\tilde{\lambda}}
\newcommand{\e}{\tilde{E}}

\title{Classification of resonance Regge trajectories and a modified Mulholland formula}
\author{D. Sokolovski*}
\affiliation{Department of Chemical Physics, University of the Basque Country, Leioa, Spain.}
\affiliation{IKERBASQUE, Basque Foundation for Science, 48011, Bilbao, Spain.}
\author{E. Akhmatskaya}
\affiliation{ Basque Center for Applied Mathematics (BCAM),\\ Building 500, Bizkaia Technology Park E-48160, Derio, Spain}
\affiliation{IKERBASQUE, Basque Foundation for Science, 48011, Bilbao, Spain.}

\begin{abstract}
{*Corresponding author: d.sokolovski@qub.ac.uk, Ph.+34 633557019}
\vspace{0.8cm}
\newline
We employ a simple potential model to analyse the effects which a Regge trajectory, correlating with a bound or a metastable state at zero angular momentum, has on an integral cross section. A straightforward modification of the Mulholland formula of Macek {\it et al} is proposed for more efficient separation of the resonance contribution.

\end{abstract}

\date{\today}

\maketitle

\vskip0.5cm
%
\section{Introduction}
There has been recent interest in cold atomic and molecular collisions where
scattering resonances may strongly influence observable cross section  (see, for example, \cite{Cost}). Earlier, Macek {\it et al}, who studied proton impact on neutral atoms \cite{Mac}, have shown that a resonance structure in the energy dependence of an integral cross section $\sigma(E)$  is simply related to the properties of Regge trajectory(ies)\cite{Conn} passing close to the real axis in the complex angular momentum (CAM) plane. The authors of \cite{Mac} also employed a method, whose origin was traced back to the work of Mulholland \cite{Mull}, to separate the resonance (pole) contribution to $\sigma(E)$ from its direct part. Regge trajectories studied in Refs.\cite{Mac} and also in a potential model employed to describe low-energy electron-atom scattering (see, for example, \cite{e}),
correlate, at zero angular momentum, $J=0$,  with bound states trapped in an attractive well.
At low energies, such trajectories tend to follow the real $J$-axis and then move into first quadrant 
of the CAM plane as $E$ increases (see also \cite{Marl}). The Mulholland formula of Ref.\cite{Mac} was extended to the multichannel case in \cite{S1} and applied to atom-diatom reactive scattering, where Regge trajectories of a different  (second) type were encountered \cite{S1},\cite{S2}. Unlike their bound state counterparts, these trajectories start in the first quadrant of the CAM plane and then, as the energy increases, approach the real $J$-axis from above. There has also been evidence  \cite{S1}-\cite{S2} that the Mulholland formula of \cite{Mac} may
underestimate the resonance contribution to $\sigma(E)$ and ascribe it to what is deemed to be a direct impact parameter-like term. The purpose of this Brief Report is to show that the trajectories of the second type are the ones which correlate with metastable rather than bound states of the 
original potential, and to suggest a straightforward modification of the Mulholland formula in order to achieve a more efficient separation of the resonance contribution. In order to do so we employ a simple potential model similar to the one used in Refs.\cite{S3}

\section{The model}
Following \cite{S3} we consider potential scattering of a unit-mass particle with an energy $E$
off a hard sphere surrounded by 
a rectangular well and
a thin semi-transparent layer, so that the effective potential of the radial Schriedinger equation, shown in Fig.1, takes the form ($\hbar=1$)
\begin{eqnarray}\label{1}
U(r) = W(r)+\Omega \delta(r-R)+(\lambda^2-1/4)/2r^2, 
\end{eqnarray}
\begin{eqnarray}\label{2}
  W(r) = \left\{
  \begin{array}{l l }
      \quad \infty & \quad \text{for $r\le R-d$}\\
    -V=const & \quad \text{for $R-d<r\le R$}\\
     \quad  0& \quad \text{for $r>R$}
  \end{array} \right.
\end{eqnarray}
\begin{figure}[ht]
\centerline{\includegraphics[width=7.5cm, angle=-0]{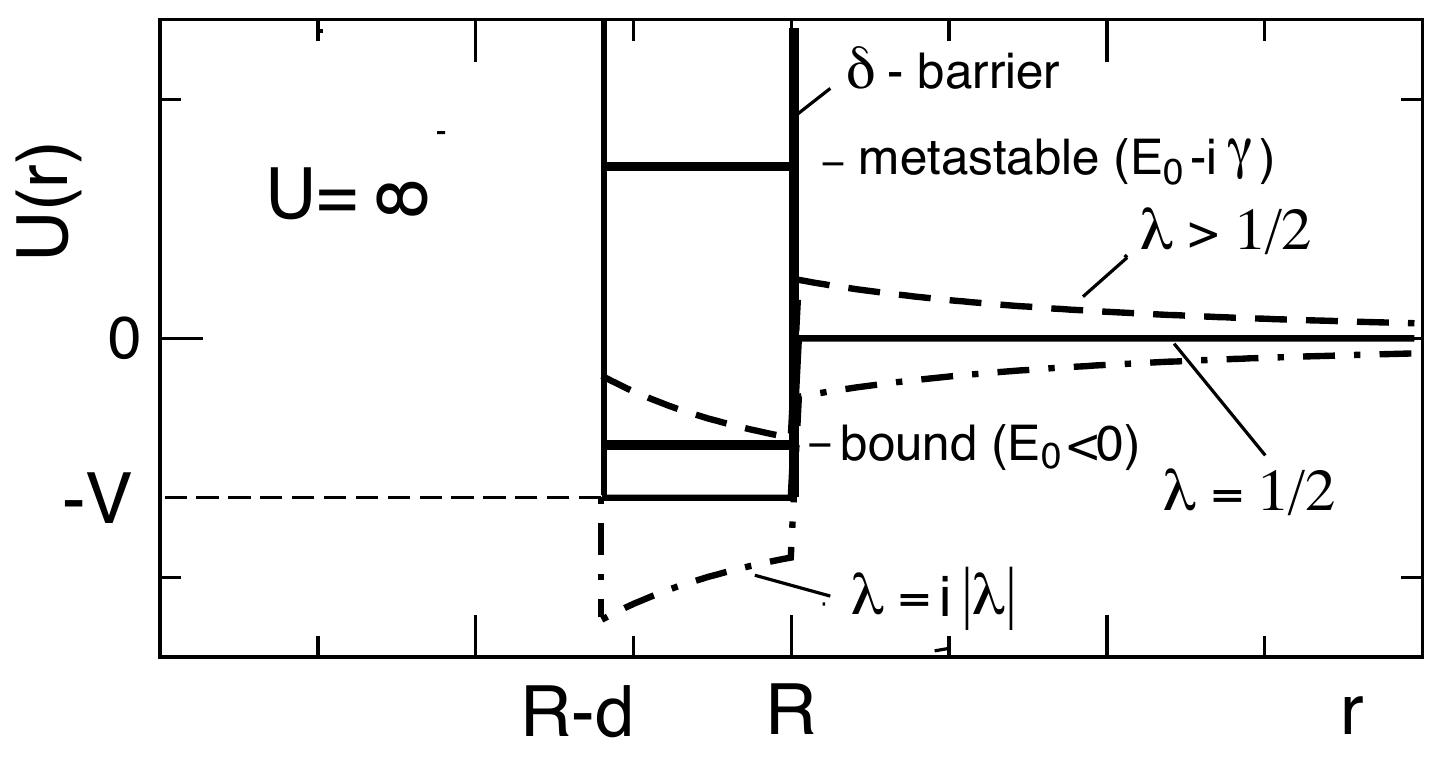}}
\caption{Effective potential $U(r)$ (a.u.) vs. $r$ for $\lambda=1/2$ (solid), $\lm>1/2$ (dashed) and
$\lm=i|\lm|$ (dot-dashed). Also shown by horizontal lines are a bound ($E_0<0$) and a 
metastable ($E_0>0$) states in the $\lm=1/2$ potential. }
\label{fig:2}
\end{figure}
where $\delta(z)$ is the Dirac delta and $\lambda$ is related to the total angular momentum $J$ as $\lambda=J+1/2$. Outside the well, for $r>R$, the wavefunction is given by
\begin{eqnarray}\label{2}
\phi(r,E,\lambda) =(\pi kr/2)^{1/2}[H^{(2)}_{\lm}(kr)+S(k,\lm) H^{(1)}_{\lm}(kr)], \q 
\end{eqnarray}
where $S(k,\lm)$ is the scattering matrix element, $k=(2E)^{1/2}$ and $H^{(1)}_{\lm}(z)$
and $H^{(2)}_{\lm}(z)$ are the Hankel functions of the first and second kind, respectively.
Inside the well, for $R-d\le r\le R$,  we have $\phi(r) =A(k,\lm) \phi_{well}(r)$,
\begin{eqnarray}\label{3}
\phi_{well}(r,E,\lm)\equiv[H^{(2)}_{\lm}(qr)- \frac{H^{(2)}_{\lm}[q(R-d)]}{H^{(1)}_{\lm}[q(R-d)]}H^{(1)}_{\lm}(qr)], \q\\
\nonumber
\end{eqnarray}
where  $q\equiv (k^2+2V)^{2}$. The values of $A$ and $S$ are determined by requiring that $\phi(r)$ be  continuous at $r=R$, while its logarithmic derivative experiences there 
a jump by $2\Omega$. Thus, we have  (prime denotes differentiation with respect to $r$)
\begin{eqnarray}\label{4}
S(k,\lm)= \Delta^{(2)}(E,\lm)/\Delta^{(1)}(E,\lm)\equiv \q\q\q\q\q\q\\
\nonumber
- \frac{([\ln H^{(2)}_{\lm}(kr)]' -[\ln \phi_{well}(r,k,\lm)]' -2\Omega}{[\ln H^{(1)}_{\lm}(kr)]' -[\ln\phi_{well}(r,k,\lm)]' -2\Omega}|_{r=R}.
\nonumber
\end{eqnarray}
 It is readily seen that 
$S(E,\lm)$ has a pole whenever $H^{(1)}_{\lm}(kr)$ alone matches onto $\phi_{well}(r,k,\lm)$ at $r=R$, i.e., provided
\begin{eqnarray}\label{5}
\Delta^{(1)}(E,\lm)=0.
\end{eqnarray}
With both $E$ and $\lm$ allowed to take complex values, Eq.(\ref{5}) defines an analytical function 
$\lt(E)$, typically single valued on a multi-sheet Riemann surface, and also its inverse, 
$\e(\lambda)$. Varying $E$ along the real $E$-axis on the Riemann sheet of interest and reading off the values $Re\lt(E)$ and
$Im\lt(E)$ one obtains a Regge trajectory and, for each value of $E$, a Regge state
$\phi(r,E,\lt(E))$. At the pole, the first term in (\ref{2}) can be neglected  and, since for $kr>>1$, $(\pi kr/2)^{1/2}H^{(1)}_{\lm}(kr)\sim \exp(ikr)$, a Regge state contains only an outgoing wave where $r$ is large.
Similarly, varying $\lm$ along the real $\lm$-axis yields a complex energy trajectory
$\{Re \e(\lm)$, $Im \e(\lm)\}$ and defines for each value of the angular momentum parameter
$\lm$ 
an outgoing wave solution (Siegert state) $\phi(r,\e(\lm),\lm)$.
\section{Two types of Regge trajectories}
 The model 
(\ref{1}) supports at least two kinds of Regge trajectories responsible for different  types of resonance structures occurring in the integral cross sections.
The two types can  be distinguished by examining the value $\e(1/2)\equiv E_0-i\gamma$
 the Siegert energy takes for  $\lambda=1/2$,  when the centrifugal potential vanishes.
\newline
{\it (I) Regge trajectory related to a bound state.}
 If  $ E_0 <0$ and $\gamma =0$, $\phi(r,\e(1/2),1/2)$ is one of the bound states
supported by the potential $W(r)+\Omega \delta(r-R)$ in Fig.1.
Since the bound state contains for $r>R$  only the decaying wave corresponding 
to the second term in Eq.(\ref{2}), it also coincides
with the Regge state $\phi(r,E_0,1/2)$. For  $E_0<E<0$, Eq.(\ref{5}) defines 
the value of $\lambda$ required to shift the bound state so that its energy aligns with $E$. 
For 
$d<<R$ the centrifugal potential in Eq.(\ref{1}) lifts the bottom of the rectangular well
and, with it, the energy of the state
approximately by 
 $(\lm^2-1/4)/[2(R-d/2)^2]$. Equating the shift to the difference $E-E_0$ yields an estimate for $\lt(E)$,
\begin{eqnarray}\label{5a}
\lt(E)^2\approx 2(E-E_0)(R-d/2)^2+1/4.
\end{eqnarray}
For $E>0$, 
 in order to generate the outgoing flux,
the centrifugal potential (and, therefore, $\lt$) acquire a positive imaginary part.  One, therefore, has 
$Im \lt(E)>0$, which also increases with the energy.
An example of such a behaviour is given in Fig.2.
\begin{figure}[ht]
\centerline{\includegraphics[width=6.5cm, angle=-0]{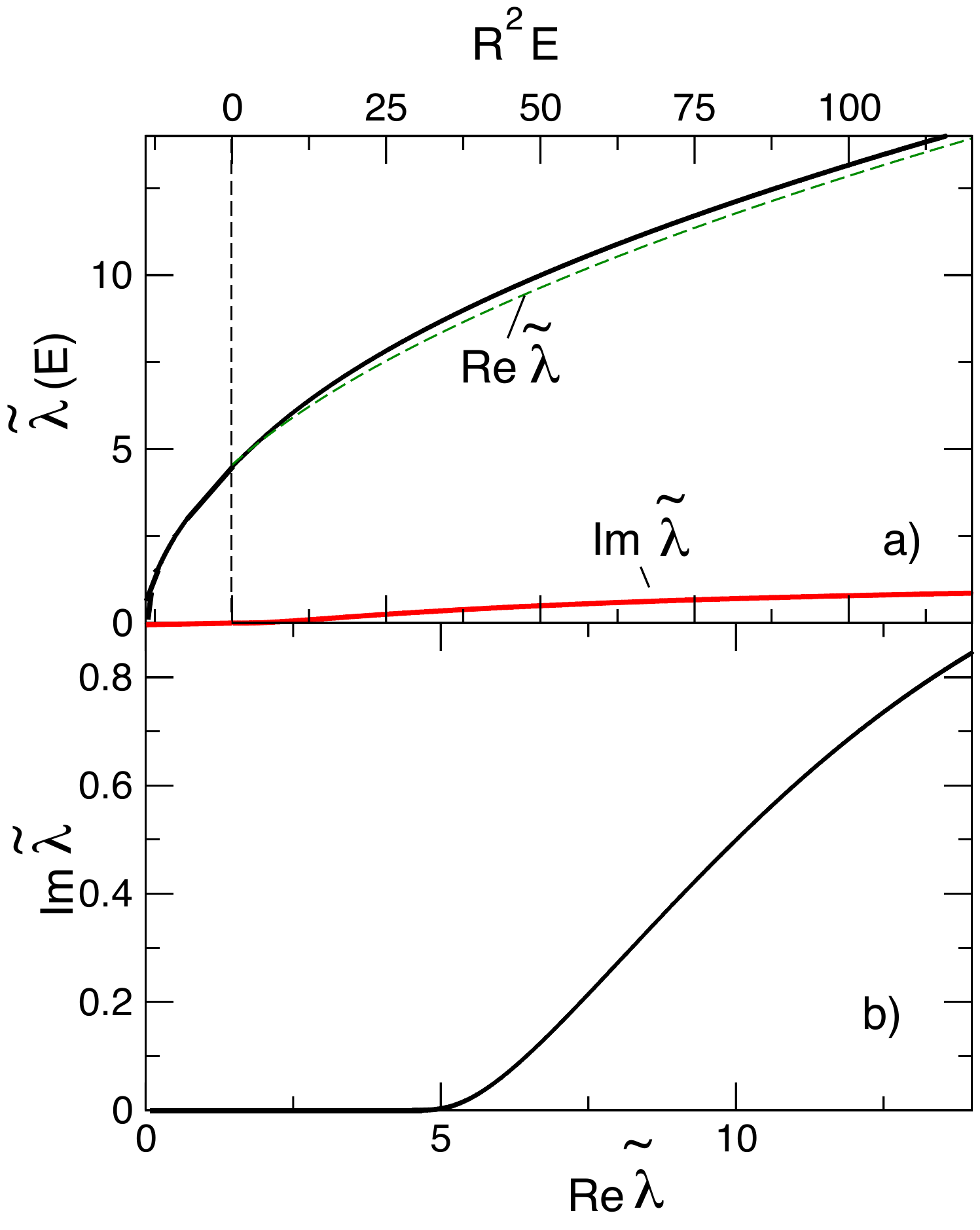}}
\caption{(color online) Regge trajectory related to the second bound state for $R^2V=165$, $\Omega R=0.5$ and $d/R=0.29$:  a) $Re\lt$ and $Im\lt$ vs. $E$. Also shown by 
the dashed line is $\lt(E)$ as given by Eq.(\ref{5a}). ; b) $Im\lt$ vs. $Re\lt$.}
\label{fig:2}
\end{figure}
Regge trajectories of this type occur in collisions between protons and neutral atoms \cite{Mac} and in 
modelling of low-energy electron-atom scattering \cite{e}.
%
 %
\newline
{\it (II) Regge trajectory related to a metastable state.} If  $E_0>0$ and  $\gamma >0$,  $\phi(r,\e(1/2),1/2)$ corresponds to a metastable state trapped 
between the hard sphere and the $\delta$-barrier (see Fig.1). It no longer coincides with the Regge 
state $\phi(r,E_0,\lt(E_0))$ but if  the resonance is long-lived, both $|\gamma|$ and $Im \lt(E_0)$
are small.
 Then the behavior of the Regge trajectory for $E$ close to $E_0$  
can be rationalised by
assuming the $\delta$-barrier impenetrable and considering the centrifugal potential required to align a positive-energy bound state with a given $E>0$.
 For $0<E<E_0$, the state must be lowered by making the well deeper and 
 the centrifugal potential attractive,
while for $E>E_0$ the state needs to be lifted by making the well shallower.
Estimated with the help of  Eq.(\ref{5a}),
 $\lt(E)$ is real for $E>E_0-(R-d/2)^2/8$ and imaginary for $E<E_0-(R-d/2)^2/8$, [c.f. Fig.3a (dashed)].
  Tunnelling across the $\delta$-barrier modifies this simple picture
 as shown in Fig.3a. Note that as $E\rightarrow 0$, $Im\lt(E)$ tends to a finite 
 value, while $Re \lt(E)\rightarrow 0$,
 and then remains zero for $E<0$, where the Regge state becomes a bound state
 in the (real valued) effective potential (\ref{1}) with an attractive centrifugal term (c.f. Fig.1).
 \begin{figure}[ht]
\centerline{\includegraphics[width=6.5cm, angle=-0]{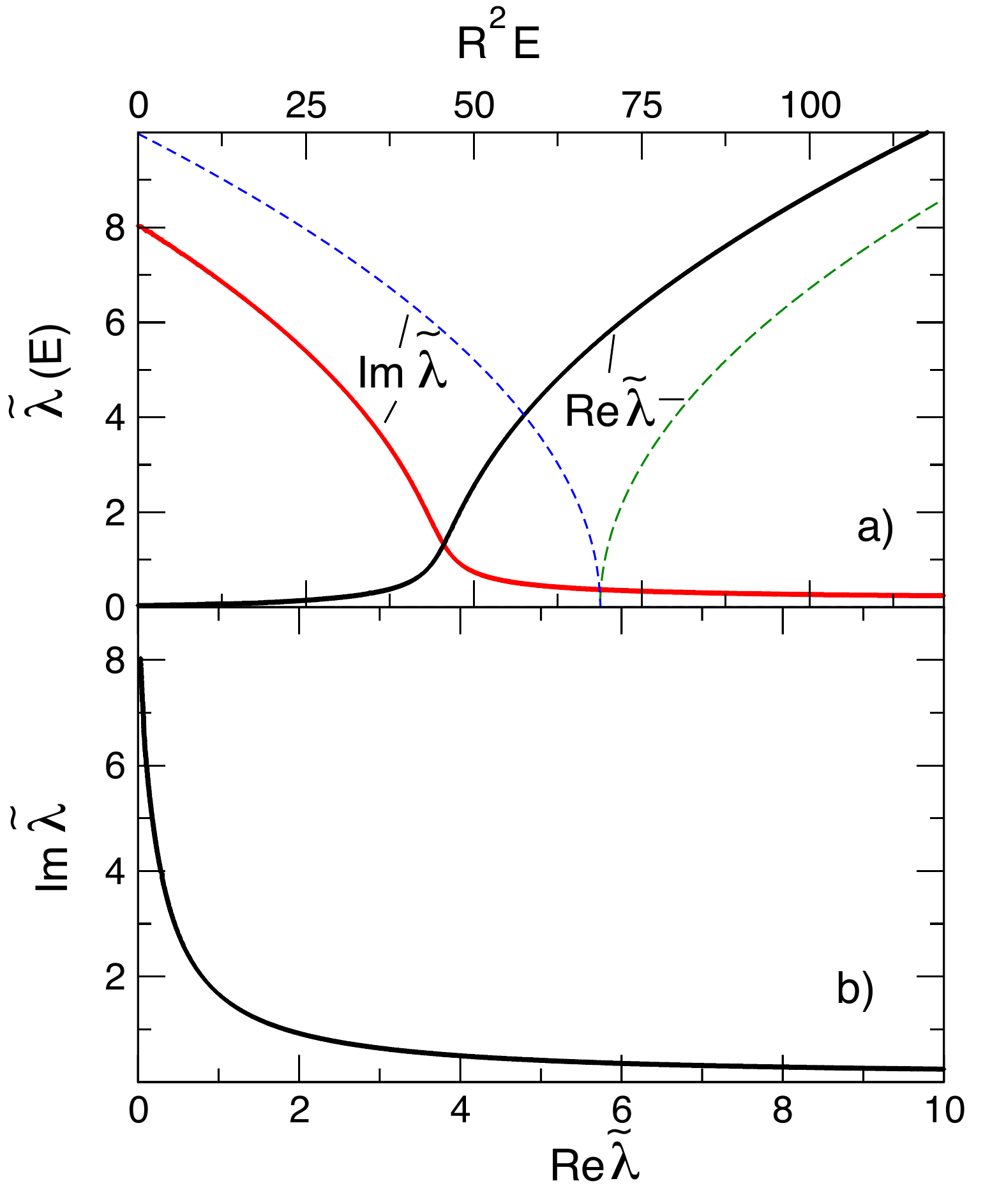}}
\caption{(color online) Regge trajectory related to the lowest metastable state for $R^2V=165$, $\Omega R=32.5$ and $d/R=0.29$:  a) $Re\lt$ and $Im\lt$ vs. $E$.
Also shown by dashed lines is the Regge trajectory for the bound state arising as  $
\Omega R\rightarrow \infty$; b) $Im\lt$ vs. $Re\lt$.}
\label{fig:2}
\end{figure}
Regge trajectories of this  second type have been found in the analysis of reactive cross sections in atom-diatom scattering \cite{S1}-\cite{S2}.
 \newline
Two kinds of Regge trajectories approach the real $\lambda$-axis in two different manners and, therefore, affect the the integral cross-sections  differently, as shown in Figs.4a and 5a.
A trajectory of type $(I)$ typically produces in a TCS a low-energy pattern. At higher energies
 the trajectory moves deeper into the CAM plane and the TCS is no longer affected by a resonance.
 A trajectory of type $(II)$ typically produces a pattern which starts at $E\approx E_0$, as the trajectory approached the real $\lm$-axis from above. Initially the pattern varies slowly with $E$, 
 with more rapid oscillations added to it at higher energies.

 \begin{figure}[ht]
\centerline{\includegraphics[width=6.8 cm, angle=-0]{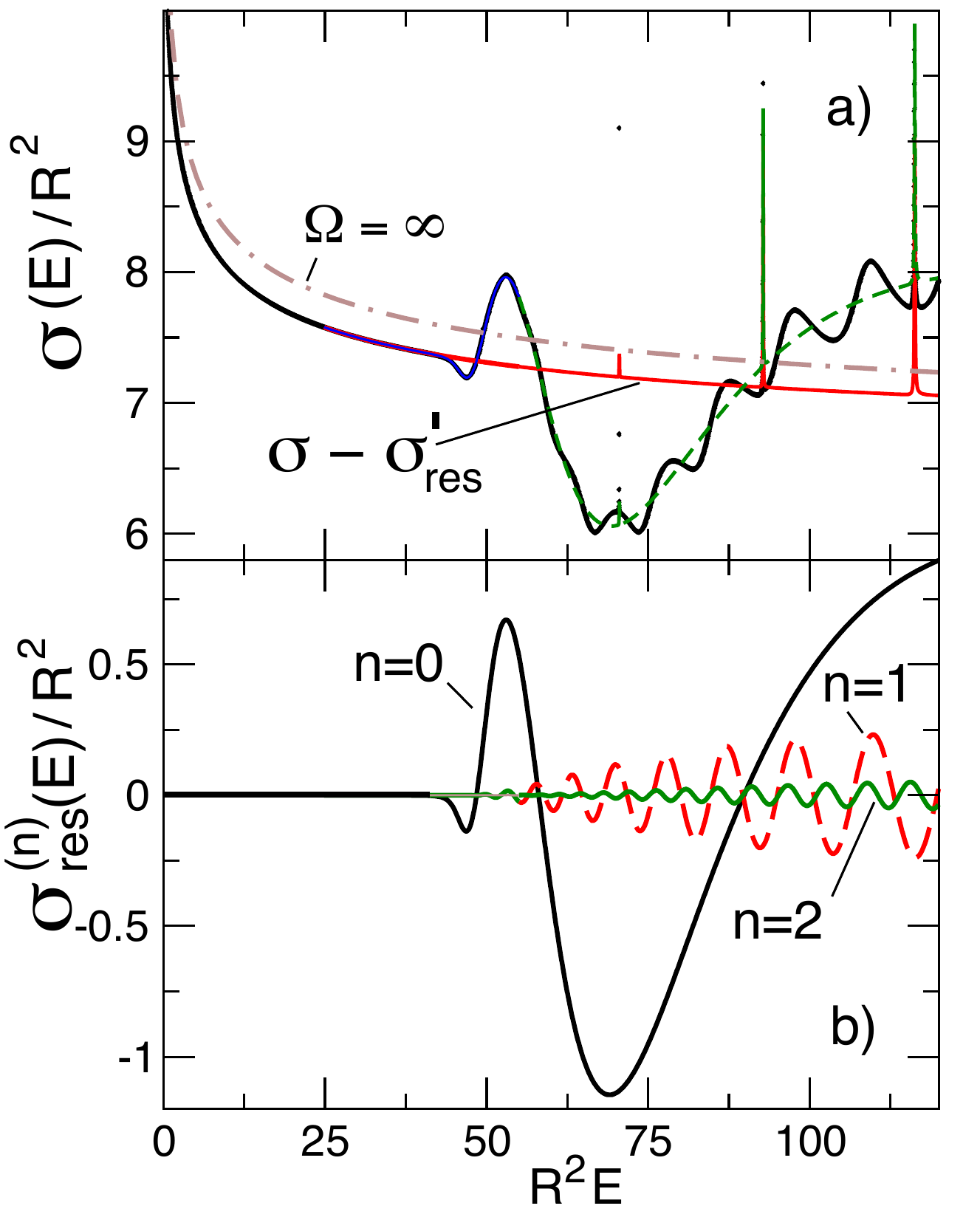}}
\caption{ (color online) Full integral cross section $\sigma(E)$ (thick solid)  and its direct part  $\sigma(E)-\sigma'_{res}(E)$ (solid)
  for the Regge trajectory in Fig.3.
Also shown are $\sigma(E)-\sigma_{res}(E)$ as given by Eq.(\ref{6}) (dashed) and 
$\sigma(E)$ for $\Omega=\infty$ (dot-dashed). Narrow peaks correspond to a Regge trajectory
related to a different bound state in the well.
 b) Mulholland contributions for $n=0,1,2$ rotations of the creeping wave around the hard-sphere core.}
\label{fig:2}
\end{figure}
 \begin{figure}[ht]
\centerline{\includegraphics[width=6.8cm, angle=-0]{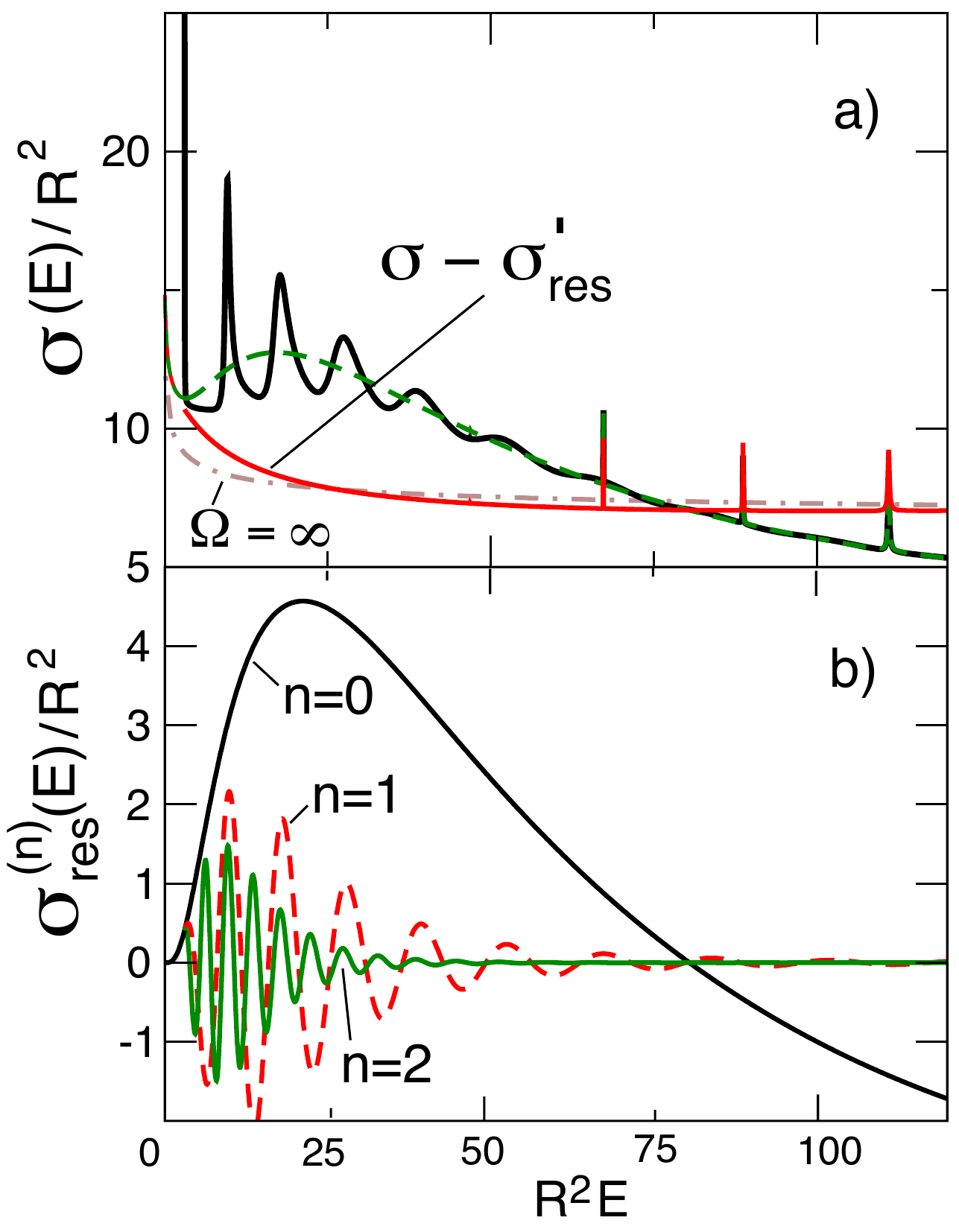}}
\caption{ (color online) Full integral cross section $\sigma(E)$ (thick solid)  and its direct part  $\sigma(E)-\sigma'_{res}(E)$ (solid)
  for the Regge trajectory in Fig.2.
Also shown are $\sigma(E)-\sigma_{res}(E)$ as given by Eq.(\ref{6}) (dashed) and 
$\sigma(E)$ for $\Omega=\infty$ (dot-dashed). Narrow peaks correspond to a Regge trajectory
related to a different bound state in the well.
 b) Mulholland contributions for $n=0,1,2$ rotations of the creeping wave around the hard-sphere core.}
\label{fig:2}
\end{figure}

\section{A modifiied Mulholland formula}
For a model (\ref{1}), the scattering amplitude contains two main contributions. The direct part
corresponds to trajectories deflected off the outer  layer represented by the $\delta$-barrier.
The resonance part corresponds to a 'creeping wave' trapped between the hard core and the outer layer. The wave, initiated by  partial waves with $\lm\approx \lt(E)$ which can penetrate the 
$\delta$-barrier, orbits the hard core and decays as the particle tunnels back into continuum.
The purpose of the CAM analysis is to decompose the integral cross-section $\sigma(E)$ into 
direct and resonance parts corresponding to the two scattering mechanisms \cite{Mac}. 
This can be achieved, for example, by using the optical theorem 
$\sigma(E)=(4\pi/k)Im f(0)$ and applying the Poisson sum formula to the partial wave expansion
of the forward scattering amplitude, $f(0)=(2ik)^{-1}\sum_{J=0}^{\infty}(2J+1)[S(k,J+1/2)-1]$  \cite{e}.  Deforming the contours of integration, one obtains the Mulholland formula \cite{Mac,e}
\begin{eqnarray}\label{6}
\sigma(E)=\sigma_1(E)+ \sigma_{res}(E) +\sigma_2(E), \q\q\q\q\q\q\q\q\q\q
\end{eqnarray}
\begin{eqnarray}\label{7}
\sigma_1(E)=4\pi k^{-2}\int_0^{\infty} Re[1-S(k,\lm)] \lm d\lm, \q\q\q\q\q\q\\
\sigma_{res}(E) \equiv \sum_{n=1}^{\infty}\sigma_{res}^{(n)}(E) =-\frac{8\pi^2} {k^{2}}Im
\frac{\lt \rho}{1+\exp(-2\pi i\lt)},\q\q\q\\
\sigma_2(E)=-\frac{8\pi^2} {k^{2}}Re \int_0^{i\infty}\frac{1-[S(k,\lm)+S(k,-\lm)]/2}{1+\exp(-2\pi i\lm)}\lm d\lm,\q\q
\end{eqnarray}
where $\rho(E) \equiv lim_{\lm\rightarrow \lt(E)}(\lm-\lt)S(k,\lm)$ is the $S$-matrix residue, and 
$\sigma_{res}^{(n)}(E)\equiv 8\pi^2 k^{-2}Im\{\lt \rho \exp[in\pi (2\lt+1)]\}$.
In Eq.(\ref{6}), the first term is assumed to yield a smooth impact parameter-type contribution
to $\sigma(E)$, $\sigma_{res}(E)$ is expected to contain the resonance effects, and 
$\sigma_{2}(E)$, which results from deformation of the integration contours, to be negligible 
or, at least, a structureless function of energy \cite{Mac}. 
\newline
The difference  
$\sigma(E)-\sigma_{res}(E)$ for the Regge trajectory in Fig.3 is shown in Fig.4a by a dashed line.
Although some of the oscillatory pattern has been removed, $\sigma(E)-\sigma_{res}(E)$, still
retains a broad structure associated with the resonance. The reason becomes clear as one notes that in Eq.(10) each $\sigma_{res}^{(n)}(E)$ contains a damping factor of $\exp(-2\pi n Im\lt)$ and, therefore, represents a contribution a creeping wave makes to the forward scattering amplitude after completing $n$ rotations around the core. With only $n>0$ terms contributing to $\sigma_{res}(E)$, the Mulholland decomposition (\ref{6}) misplaces the first  ($n=0$) contribution
by including it in the direct term $\sigma_1$. This can be remedied by subtracting $\sigma_{res}^{(0)}(E)$ shown in Fig.4b
 from $\sigma_1$ and  adding it to $\sigma_{res}$ so that the modified Mulholland formula reads
 $\sigma=\sigma'_1+ \sigma'_{res} +\sigma_2$ with 
\begin{eqnarray}\label{8}
\sigma'_1(E)=4\pi k^{-2}\int_{\Gamma}Re[1-S(k,\lm)] \lm d\lm \q\q\q\q\q\q\q\q\q\\
\nonumber
\sigma'_{res}(E) \equiv \sum_{n=0}^{\infty}\sigma_{res}^{(n)}(E) =\frac{8\pi^2} {k^{2}}Im
\frac{\lt \rho}{1+\exp(2\pi i\lt)},\q\q\q
\end{eqnarray}
where contour $\Gamma$ starts at the origin, ends at $+\infty$ and passes above the Regge pole
at $\lt$ in the first quadrant of the complex $\lambda$-plane. The difference $\sigma(E)-\sigma'_{res}(E)$ shown in Fig.4a (solid) contains no resonance structure and closely resembles the integral cross section produced by the direct trajectories reflected off an impenetrable sphere of radius $R$ (dot-dashed).
\newline Results of the same analysis for the bound state Regge trajectory in Fig.2 are shown 
in Fig.5. In a similar way, the modified Mulholland decomposition (\ref{8}) allows to separate effects
of resonance capture from direct scattering off the outer layer,
while the original formula (\ref{6}) underestimates the resonance contribution.
 Figure 5b shows partial resonance cross sections $\sigma_{res}^{(n)}(E)$ for $n=0,1,2$. Note that in Figs. 4b and 5b  $\sigma_{res}^{(n)}(E)$ with $n>0$ are concentrated in the regions where a Regge trajectory passes close to  real $\lambda$ axis and produces sharp features in $\sigma(E)$.
 \section{Conclusions}
In summary, the types of behaviour shown in Figs.2 and 3 are typical of Regge trajectories
which correlate with a bound and a metastable state in the original ($J=0$) potential, respectively.
Both can be understood in terms of  a centrifugal potential required to align a resonance state with
a given energy $E$. In both cases, a straightforward modification of the Mulholland formula allows for
a more efficient separation of the resonance and the direct contributions. 
Although a simple model was used for illustrational purposes, we expect these conclusions to be valid in a more general case.
Extension to inelastic 
and reactive scattering will be considered elsewhere.
 \begin{acknowledgements}
 This work was supported by the Basque Goverment grant IT472
and MICINN (Ministerio de Ciencia e Innovacion) grant FIS2009-12773-C02-01.
\end{acknowledgements}

\end{document}